\documentclass[english]{article}
\usepackage[utf8]{inputenc}
\usepackage[english]{babel}
\usepackage{xcolor}
\usepackage[normalem]{ulem}
\usepackage{geometry}
\usepackage{float}
\usepackage{amsmath}
\usepackage{graphicx}
\usepackage{natbib}
\usepackage{lipsum}
\floatstyle{ruled}
\newfloat{algorithm}{tbp}{loa}
\providecommand{\algorithmname}{Algorithm}
\floatname{algorithm}{\protect\algorithmname}

\newcommand{\lyxaddress}[1]{
	\par {\raggedright #1
	\vspace{1.4em}
	\noindent\par}
}

\begin{document}

\title{Interplanetary Shock-induced Magnetopause \\ Motion: Comparison between Theory and Global Magnetohydrodynamic Simulations}

% The Dream Team
\author{{R. T. Desai$^1$\thanks{corresponding author: ravindra.desai@imperial.ac.uk}} , M. P. Freeman$^2$, J. P. Eastwood$^1$, J. W. B. Eggington$^1$,  \\ M. O. Archer$^1$ Y. Y. Shprits$^{3,4}$, N. P. Meredith$^2$, F. A. Staples$^5$, I. J. Rae$^6$, \\  H. Hietala$^1$, L. Mejnertsen$^1$, J. P. Chittenden$^1$, R. B. Horne$^2$}

\date{}
\maketitle

\vspace{-2em}
\lyxaddress{\begin{center}
$^1$Blackett Laboratory, Department of Physics, Imperial College London, London, UK
\par\end{center}}
\vspace{-3em}
\lyxaddress{\begin{center}
$^2$British Antarctic Survey, Cambridge, UK
\par\end{center}}
\vspace{-3em}
\lyxaddress{\begin{center}
$^3$GFZ German Research Centre for Geosciences, Potsdam, Germany
\par\end{center}}
\vspace{-3em}
\lyxaddress{\begin{center}
$^4$Department of Earth, Planetary, and Space Sciences, University of California, California, USA
\par\end{center}}
\vspace{-3em}
\lyxaddress{\begin{center}
$^5$Mullard Space Science Laboratory, University College London, Surrey, UK
\par\end{center}}
\vspace{-3em}
\lyxaddress{\begin{center}
$^6$Department of Mathematics, Physics and Electrical Engineering, Northumbria University, Newcastle Upon Tyne, UK
\par\end{center}}
\vspace{-2em}

\begin{abstract}

The magnetopause marks the outer edge of the Earth's magnetosphere and a distinct boundary between solar wind and magnetospheric plasma populations.
In this letter, we use global magnetohydrodynamic simulations to examine the response of the terrestrial magnetopause to fast-forward interplanetary shocks of various strengths and compare to theoretical predictions. 
{The theory and simulations indicate} the magnetopause response can be characterised by three distinct phases; an initial acceleration as inertial forces are overcome, a rapid compressive phase comprising the majority of the distance travelled, and large-scale damped oscillations with amplitudes of the order of an Earth radius. The two approaches agree in predicting subsolar magnetopause oscillations with frequencies 2--13 mHz {but the simulations notably predict larger amplitudes and weaker damping rates}.
This phenomenon is of high relevance to space weather forecasting {and provides a possible explanation for magnetopause oscillations observed following the large interplanetary shocks of August 1972 and March 1991}.
\end{abstract}

\section{Introduction}

The Earth's magnetopause exists in a delicate balance between forces exerted between the impinging solar wind and the Earth's intrinsic magnetic field. The subsolar magnetopause is typically located approximately ten Earth radii (R$_E$) upstream but, during periods of enhanced solar wind forcing, this can be compressed to half this distance and inside the drift paths of radiation belt electrons and protons \citep{Shprits06} and the orbits of geosynchronous satellites \citep{Cahill99}. 
Moreover, magnetopause motion can drive gloabl ultra-low-frequency (ULF) pulsations \citep{Li97,Green04} and intense ionospheric and ground induced current systems \citep{Fujita03,Smith19}.  
The dynamics and location of the magnetopause are therefore of wide relevance to the understanding of planetary magnetospheres and to space weather forecasting.

{The location and shape of the magnetopause was initially theoretically predicted to depend on the pressure exerted by a stream of charged particles from the Sun \citep{Chapman1931} and its three dimensional geometry was derived based on solar wind dynamic pressure alone \citep{Mead64}. Measurements with in-situ spacecraft broadly confirmed these predictions and were then used to derive a large suite of empirical models of the magnetopause location \citep[e.g.][and references therein]{Shue1998} based on elliptical and parabolic functions.  These empirical studies revealed additional influences from the Interplanetary Magnetic Field (IMF) orientation, which modulates magnetic reconnection and the \citet{Dungey1961} cycle, solar wind magnetic pressure and dipole tilt \citep{Lin2010}, IMF cone angle \citep{Merka2003}, and ionospheric conductivity and solar wind velocity \citep{Nemecek2016}}. These best-fit models are, however, static and can deviate when compared to specific observations \citep{Samsonov2019}, particularly during extreme solar wind conditions with discrepancies of $>$1 R$_E$ observed when located less than 8 R$_E$ upstream \citep{Staples2020}.

Satellite observations have revealed that the magnetopause boundary exists in a perpetual state of motion \citep{Bowe90}. Solar wind pressure variations drive the magnetopause response which results in fast magnetosonic waves that can couple to poloidal and toroidal Alfv\'en modes of the large-scale magnetospheric fields \citep{Southward1974,Kivelson1984}.
Bow shock- and magnetosheath-generated phenomena, including; hot flow anomalies \citep{Burgess1989}, magnetosheath jets \citep{Hietala09}, foreshock cavities \citep{Sibeck2002} and bubbles \citep{Omidi2010}, similarly produce pressure fluctuations which elicit magnetopause motion.

Only four studies have formally examined the directly driven response of the magnetopause to upstream pressure variations. \citet{Smit1968} initially formulated magnetopause motion as a simple harmonic oscillator consisting of inertial, damping, and restoring forces. \citet{Freeman1995} and \citet{Freeman1998} subsequently used the Newton-Busemann approximation to develop a formal consistent theory of the magnetopause as an elastic membrane which could be applied locally. \citet{Borve2011} similarly modelled the magnetopause response to solar wind pressure pulses and found qualitative agreement with 2--D MHD simulations. \citet{Freeman1995} and \citet{Borve2011} notably predict magnetopause oscillations to be strongly damped.

These studies, however, focussed on small perturbations in solar wind dynamic pressure. Fast-forward inter-planetary (IP) shocks, as occur at the front of Interplanetary Coronal Mass Ejections and corotating-interaction-regions, can rapidly compress the magnetosphere in just a few minutes \citep{Smith76,Araki94}, and present a {further} regime for studying magnetopause motion. Global magnetohydrodyanmic (MHD) codes are able to self-consistently model the dynamic solar wind-magnetosphere interaction for a wide variety of solar wind conditions and, in this study, we test theoretical predictions using global MHD simulations to constrain nonlinear magnetopause behaviour across extreme scenarios for which spacecraft observations are limited or unavailable. 

This letter is organised as follows: Section 2 describes the Gorgon Global MHD model, the simulation parameters including the IP shocks considered and theory of the magnetopause. Section 3 then describes the simulations conducted and the comparison to theory. Sections 4 concludes with a summary discussed in relation to space weather forecasting.

\section{Method}

\subsection{Global-MHD}

Gorgon is a 3-D simulation code with resistive MHD and hydrodynamic capabilities, originally developed to study high-energy-density laboratory plasmas \citep{Chittenden2004,Ciardo2007}. Gorgon has been adapted and applied to planetary magnetospheres in several contexts, including: the inclined and rotating Neptunian magnetosphere \citep{Mejnertsen2016}, the variable motion of the terrestrial bow shock \citep{Mejnertsen2018}, and the effects of dipole-tilt on terrestrial magnetopause reconnection and ionospheric current systems \citep{Eggington2020}. 

The MHD equations are implemented to represent a fully ionised quasi-neutral hydrogen plasma on a 3-D uniform Eulerian cartesian grid. A second order finite volume Van Leer advection scheme uses a vector potential representation of the magnetic field on a staggered \citet{Yee1966} grid which maintains a divergence free magnetic field to machine-precision. The system is closed assuming an ideal gas and stepped forward with a variable time-step using a second order Runge Kutta scheme. These numerics conserve the internal energy, rather than total energy, which negates negative pressures. A split magnetic field is implemented \citep{Tanaka1994} where the curl-free dipole-component is omitted from the induction equation which reduces discretisation errors within the magnetosphere. 
 A \citet{Boris1970} correction is used to limit the Alfv\'en speed in the presence of a reduced speed of light, and a Von Neumann artificial viscosity is applied to accurately capture shock physics and improve energy conservation \citep{Benson1992}. 
Due to its heritage in simulating laboratory plasmas, Gorgon also includes individual pressure terms for protons and electrons, Ohmic heating based upon the Spitzer resistivity, optically thin radiative loss terms, and electron-proton energy exchange. These are, however, vanishingly small within collisionless magnetospheric plasmas. Magnetic reconnection therefore develops through numerical diffusion alone.

The simulation domain extends from -20 to 100 R$_E$ in X and -40 to 40 in Y and Z, with a uniform grid spacing of 1/2 R$_E$, and which corresponds to GSM coordinates with --X. 
An inflow boundary condition is located on the sun-ward edge (--X) where the solar wind propagates into the domain, and outflow boundary conditions are used at the tailward X, and Y and Z boundaries. The dipole is located at the origin and the inner ionospheric boundary is located at $\mid$3$\mid$ R$_E$ with a 370 cm$^{-3}$ fixed density of cold 0.1 eV plasma which diffuses outward to form a rudimentary plasmasphere. The ionosphere at the inner boundary \citep{Eggington2018} is represented by a thin conducting shell, upon which the generalized Ohm's law is solved for a given ionospheric conductance profile to obtain an electrostatic potential \citep{Ridley2004}. The corresponding electric field then modifies the plasma flow via the associated drift velocity. 
The simulation is initialised with a dipole field with an exponentially decreasing low plasma density through the domain and with a mirror dipole within the solar wind to produce a B$_x$=0 surface \citep{Raeder03}. Constant solar wind conditions of n$_0$ = 5 cm$^{-3}$, B$_z$ = -2 nT, T$_i$ = T$_e$ = 5 eV, v$_x$ = 400 km s$^{-1}$, as shown in Table 1, are run for two hours with geomagnetic dipole moment M$_z$ = 7.94$\cdot$10$^{22}$ Am$^2$ to produce a fully formed magnetosphere.

\subsection{Interplanetary Shocks}

{Interplanetary shocks are produced at the interface of plasma regimes in the solar wind when the relative speed of the shock structure to the ambient solar wind exceeds the magnetosonic velocity} \citep{Kennel85}. 
Fast-forward shocks are characterised by an increase in velocity, density, pressure and magnetic field strength, as produced at the leading edge of impulsive phenomena such as interplanetary coronal mass ejections \citep{Burlaga71} and between fast and slow solar wind streams as these boundaries steepen into corotating interaction region-driven shocks \citep{Smith76}.
\vspace{-2em}
\begin{center}
\begin{table}[ht]
\centering
\caption{Rankine-Hugoniot jump conditions for four fast-forward perpendicular IP shocks corresponding to four Gorgon simulations with the same initial solar wind conditions. }
\centering
\begin{tabular}{|c|c|c|c|c|c|c|}
\hline
\textbf{}           & \textbf{\begin{tabular}[c]{@{}c@{}}n\\ {[}cm$^{-3}${]}\end{tabular}} & \textbf{\begin{tabular}[c]{@{}c@{}}v$_x$ \\ {[}km s$^{-1}${]}\end{tabular}} & \textbf{\begin{tabular}[c]{@{}c@{}}D$_p$ \\ {[}nPa{]}\end{tabular}} & \textbf{\begin{tabular}[c]{@{}c@{}}T\\ {[}eV{]}\end{tabular}} & \textbf{\begin{tabular}[c]{@{}c@{}}B\\ {[}nT{]}\end{tabular}} & \textbf{\begin{tabular}[c]{@{}c@{}}v$_{shock}$\\ {[}km s$^ {-1}${]}\end{tabular}}\\
 \hline
\textbf{Solar Wind} & 5                                                                                            & 400                               & 1.34                                                              & 5.0                                                                     & [0, 0, -2]                                                               & -                                                                         \\ \hline
\textbf{Shock I}    & 7.5                                                                                            & 500                               & 3.14                                                                & 210.1                                                                   & [0, 0, -3]                                                               & 700                                                                                                 \\ \hline
\textbf{Shock II}    & 10                                                                                            & 600                               & 6.03                                                                & 416.3                                                                   & [0, 0, -4]                                                               & 800                                                                                                \\ \hline
\textbf{Shock III}    & 15                                                                                            & 800                            & 16.1                                                                   & 830.1                                                                  & [0, 0, -6]                                                               & 1000                                                                                                \\ \hline
\textbf{Shock IV}    & 20                                                                                            & 1000                             & 33.5                                                                 & 1244.3                                                                 & [0, 0, -8]                                                               & 1200                                                             \\ \hline
\end{tabular}
\end{table}
\label{tableshocks}
\end{center}
\vspace{-1em}

Four perpendicular fast-forward shocks of varying strengths are injected into the solar wind within four separate Gorgon simulations in order to characterise the magnetospheric response to impulsive events of varying magnitude. {Perpendicular shocks denote shock geometries where {the magnetic field is orthogonal to the shock normal}. The jump in solar wind conditions therefore manifests as a spatially uniform front.} The shocks are calculated in accordance with the Rankine-Hugoniot conditions {\citep{Priest14}} with the four jumps from the same initial solar wind, as shown in Table 1. Shock I shows a modest jump in all parameters representative of the median southward IP shock properties pbserved at 1 au during solar minimum \citep{Echer03}. The solar wind number density, n, jumps from 5 to 7.5 cm$^{-3}$, southward IMF, B$_z$, from -2 to -3 nT and solar wind velocity, v$_x$, from 400 to 500 km s$^{-1}$. Shocks II, III and IV  represent increasingly stronger cases up to the maximum possible four-fold increase in the solar wind density and magnetic field for Shock IV with a solar wind velocity jump of 400 to 1000 km s$^{-1}$ . 
All shocks are travelling at 200 km s$^{-1}$ in the solar wind frame which is at the upper bound of the 50--200 km s$^{-1}$ range {typically observed at 1 au \citep[e.g.][]{Berdichevsky00}}.

The parameters of Shock IV are judged as an estimate \citep{Hudson97} of the extreme IP shock of 24 March 1991 which rapidly compressed the magnetosphere over the course of minutes and promptly formed a new radiation belt in the slot region  \citep{Blake1992,Horne15}.
Space weather events of this extremity are rare \citep{Riley12,Meredith17}, but not unique as there are other examples where the magnetopause has been observed inside geosynchronous orbit as low as 5.2 R$_E$.  \citep{Cahill77}. It is also important to note that greater shock velocities of over twice that of Shock IV are possible. For example, on 23 July 2012, the STEREO-A spacecraft observed a non-Earth directed fast-forward shock with a velocity of $\approx$ 2250 km s$^{-1}$ \citep{Russell2013} and theoretical studies have highlighted the possibility of shock velocities over 3,000 km/s emerging from the solar corona \citep{Yashiro04,Gopalswamy05,Tsurutani14} with corresponding velocities of up to $\approx$ 2,750 km s$^{-1}$ manifesting at 1 AU \citep{Desai20}.

\subsection{Theory}
\label{SHM}

To understand the motion of the subsolar magnetopause in response to an IP shock, it is useful to consider the forces acting upon it. The following is based on the theory of \citet{Freeman1995} and \citet{Freeman1998}, and references therein. In steady state, the geocentric distance to the subsolar magnetopause, R, is well approximated by a balance between the pressure exerted on the magnetopause by the shocked solar wind and the magnetic pressure of the compressed dipole magnetic field of the Earth
\begin{equation}
    s \rho u^2 = \frac{f^2B^2_e R^6_E}{2 \mu_0 R^6},
    \label{numerical}
\end{equation}
where $\rho$ and $u$ are the solar wind density and speed, respectively, and $s$ = 1 in the Newtonian approximation. B$_{eq}$ = 31100 nT is the equatorial magnetic field strength at $1$ R$_E$, and $\mu_0$ is the permeability of free space. $f \approx 2.44$ is the typical dipole compression factor but this can theoretically vary between $f = 2$ for a plane magnetopause to $f = 3$ for a spherical magnetopause {\citep{Mead64}}. 

\begin{equation}
    m \frac{d^2 R}{d t^2} = \frac{f^2B^2_{eq} R^6_E}{2 \mu_0 R^6} - s \rho (u_\infty+\frac{dR}{dt})^2,
    \label{numerical2}
\end{equation}
where the final term is the Newtonian pressure applied to the now-moving magnetopause and the subscript $\infty$ denotes the constant post-shock solar wind values. The inertial mass m is expected to be that of the subsolar magnetosheath column. Writing m = c $\rho_{\infty}$ R$_{\infty}$, where R$_{\infty}$ is the final equilibrium position, we estimate c $\approx$ 1.2 in this case. Also rewriting the magnetic pressure term using the final equilibrium version of Equation \ref{numerical}, Equation \ref{numerical2} becomes
\begin{equation}
    \frac{d^2 R}{d t^2} + \frac{s}{c R_\infty} \left[ \left( u_\infty + \frac{dR}{dt} \right) ^2 - u_\infty^2 \left(\frac{R_\infty}{R}\right)^2 \right] = 0
    \label{numerical3}.
\end{equation}
Linearising Equation \ref{numerical3} by substituting R(t) = R$_\infty$ + r(t), assuming r $<<$ R$_\infty$, and retaining only first-order terms, the equation of motion becomes:
\begin{equation}
    \frac{d^2 r}{d t^2} + \left( \frac{2}{K \tau} \right)  \frac{d r}{d t} + \left( \frac{6}{K \tau^2} \right) r = 0
 \label{linear}
\end{equation}
where $\tau$ = R$_\infty$ / u$_\infty$ is the characteristic system time scale, and K = c/s. The homogeneous second-order ordinary differential Equation \ref{linear} is that of a damped simple harmonic oscillator whose solution is an exponentially-decaying sinusoid, 
\begin{equation}
    r = A e^{-bt}cos(\omega t + \phi),
 \label{linear2}
\end{equation}
where b = 1 / (K $\tau$) and $\omega$ = b $\sqrt(6K-1)$. For a stationary pre-shock magnetopause at position, R$_0$, we have tan($\phi$) = -b / $\omega$ and A cos($\phi$)  = R$_0$ - R$_\infty$.

\section{Results} \label{23july2012}

\subsection{Shock-Magnetosphere Interaction}

\begin{figure*}[ht]
%\centering
\hspace*{-2cm}
\includegraphics[width=1.25\textwidth]{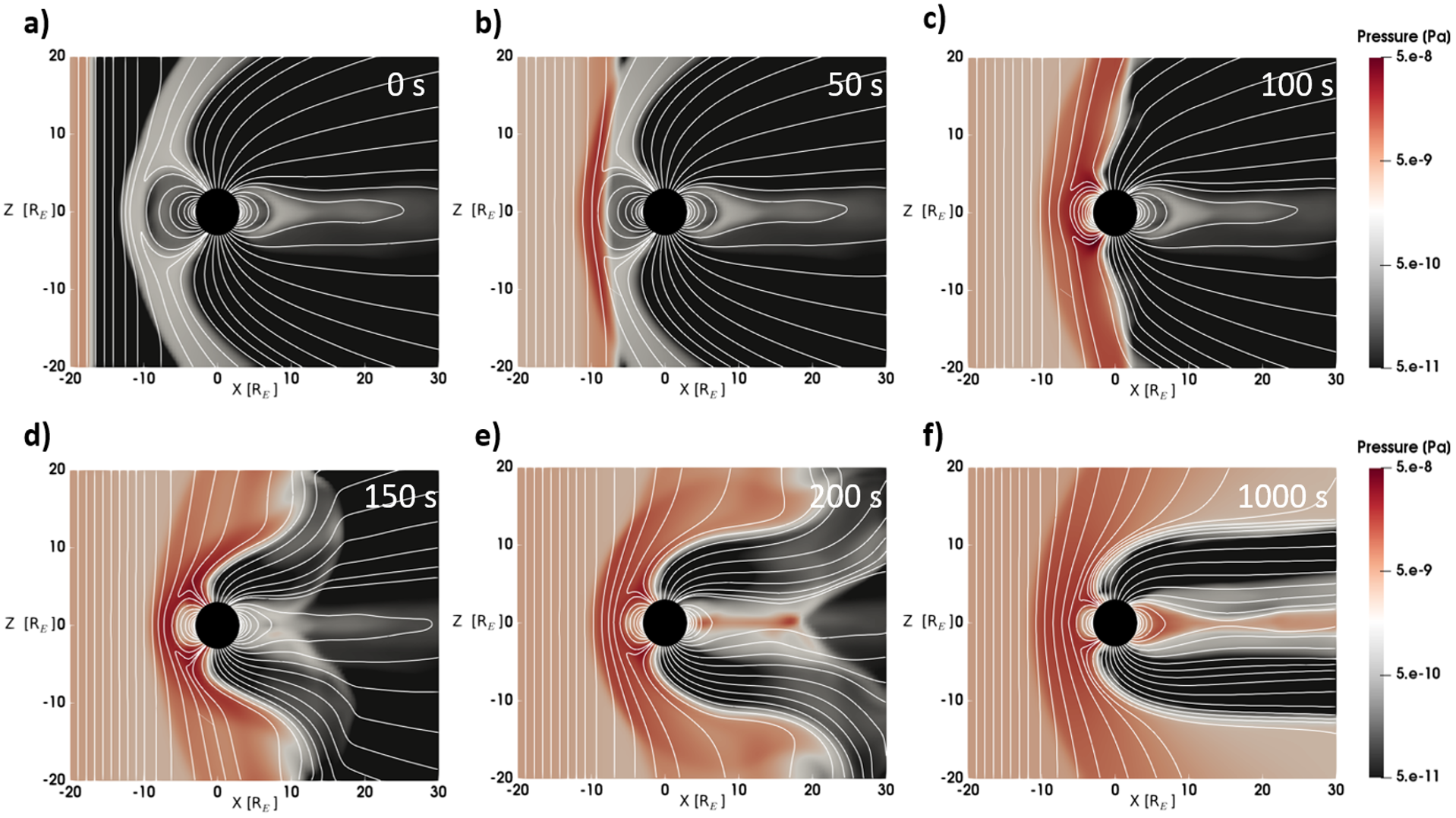}
\caption{Gorgon pressure in the x--y plane at six instances corresponding to before, during, and after, IP Shock IV impacts the simulated magnetosphere. Selected magnetic field lines are depicted in white and the shock parameters are listed in Table 1.  
\label{compression}}
\end{figure*}

Figure \ref{compression} shows the Gorgon pressure at six stages during the simulation of Shock IV, starting within the upstream solar wind, then at four stages within the magnetosphere, and then sometime after when the system has reached a new compressed steady state. Selected magnetic field lines are depicted in white and the shock moves through the domain shown in just over 200 seconds. 

The IP shock slows down upon passing through the bow shock and panel (b) shows it develops a curved front as it propagates through the dense magnetosheath \citep{Samsonov06,Andreeova11}. 
The subsequent impact on the magnetopause disrupts the pressure-balanced equilibrium which initiates the commencement phase associated with geomagnetic storms \citep{Smith86,Araki94}.
The initial magnetospheric state shows pressures below 1 nPa and the enhanced solar wind pressure consequently produces magnetosheath pressures over an order of magnitude higher. The tailward propagating magnetosonic pulse, panels (d--e), subsequently produces enhanced plasma sheet pressures, thinning of the tail current sheet and induces near-Earth tail reconnection {\citep{Oliveira14}}. The enhanced dynamic pressure in the solar wind compresses the magnetopause boundary from its initial position near --10 R$_E$ to its final position near --6 R$_E$.

\subsection{Subsolar Magnetopause}

The magnetopause can be characterised as possessing a finite thickness from several ion gyroradii of several hundred kilometres \citep{Le94} to over half an Earth radius \citep{Kaufmann73}. The different plasma conditions on either side results in this being an asymmetric structure. 
Along the sub-solar line the shocked solar wind first slows and diverts at the fluopause \citep{Palmroth03} which, based on the gradient of the velocity stream lines, is initially determined at $\approx$ --10.9 R$_E$. The southward oriented magnetic field then passes through zero at --10.75 R$_E$ as it tends to the significantly larger positive magnetospheric fields. Further inward, the peak in the magnetopause current density is located at --10.45 R$_E$, which is then followed by a local depletion in the plasma density at --10.1 R$_E$. 
In this study we determine the magnetopause position using the B$_z$ = 0 condition, which provides a
consistent measure for southward IMF regardless of solar wind conditions and stand-off distance.

\begin{figure*}[ht]
\centering
\hspace*{-1cm}
\includegraphics[width=1.1\textwidth]{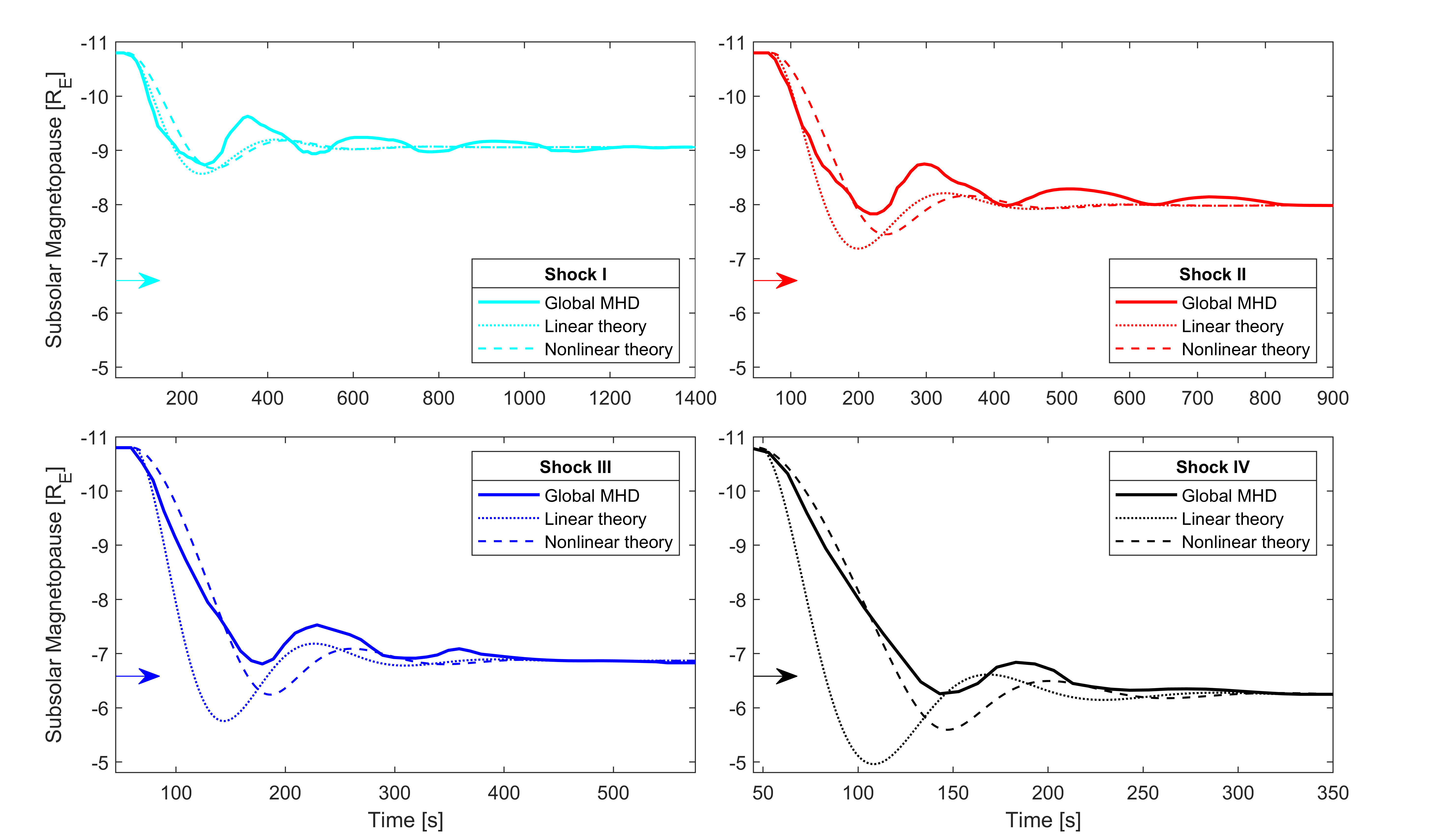}
\caption{Simulated subsolar magnetopause stand-off distance compared to linear and nonlinear theoretical predictions for the four Shocks listed in Table 1. For reference the location of geosynchronous orbit is annotated with an arrow.
\label{subsolar}}
\end{figure*}

Figure \ref{subsolar} shows traces of the Gorgon subsolar magnetopause stand-off distances (solid lines) over time for the four shocks simulated. The motion of the magnetopause appears as three distinct phases. The first involves an acceleration as the inertia of the magnetosheath is overcome. The second appears as a rapid compressive phase which comprises the majority of the change in standoff distance.
The end of this rapid compression marks the third stage of large-scale oscillatory motion with amplitudes of the order of an Earth radius before the magnetopause reaches pressure-balanced equilibrium. 

Shock I has the smallest compressive phase as the final oscillations around pressure balance appear of a comparable magnitude to the total stand-off distance travelled. For increasing shock strengths, the duration of the compressive phase increases and the amplitudes and also frequencies of the oscillations appear to decrease. 
The underlying position about which the oscillations occur shifts Earthward as the oscillations are damped away which may be attributed to changing conditions within the sheath, see Figure \ref{compression}. 
The oscillations also appear more strongly damped for the stronger shocks with Shock IV producing magnetopause oscillations for approximately 300 seconds compared to Shock I which produces oscillations which last four times as long. 
Shocks I and II also feature more oscillations than III and IV, indicating that they have a weaker damping ratio.

Also shown in Figure \ref{subsolar} are the subsolar magnetopause motions of the four shocks predicted by the nonlinear numerical solution of Equation \ref{numerical3} (dashed lines) and the linear solution given by Equation \ref{linear2} (dotted lines), using c=1.2, s=1, and values for $\rho_\infty$, u$_\infty$, R$_0$, and R$_\infty$ taken from the simulations. As expected, the linear solution is most similar to the nonlinear numerical solution for Shock I, where the approximation r$<<$R$_\infty$ is most valid. The difference increases with shock strength, especially in the initial phase when the second-order (dR/dt)$^2$ term in Equation \ref{numerical3} is not negligible. Nevertheless, the linear theory is instructive in explaining the qualitative response characteristics of a finite magnetopause response time, overshoot, and decaying oscillation. 
For the nonlinear theory solution, the oscillation period of the simulation {produces good agreement} in all cases but the initial response time and oscillation damping rate are both progressively overestimated compared to the simulation with weakening shock strength. This suggests that the second term in Equation \ref{numerical3} may be an oversimplification in the weak shock limit, and particularly the (dR/dt)$^2$ term within it because the linear theory that neglects (dR/dt)$^2$ actually captures the initial simulation response better than the nonlinear theory for Shocks I and II. It should also be noted that the initial response is very sensitive to the initial conditions in the magnetosheath (not shown) which may differ in the simulation from those assumed in the theory. 

The linear theory is instructive in understanding the underlying physics of the magnetopause response. {The nonlinear theoretical solutions of Equations \ref{numerical3} provide a means to extend this to larger peturbations} but the solutions are {still} necessarily dependent on the choice of coefficients and the assumptions behind these such as the shape of the magnetopause surface and constant sheath thickness. 
Further effects such as magnetic reconnection, magnetosheath heating, finite solar wind mach numbers and wave speeds, {and the reflection at the pulse of the inner boundary back onto the magnetopause \citep{Li1993,Samsonov07}}, are also not accounted for. 
The time-dependent and self-consistent numerical solutions to the MHD equations, as solved by Gorgon, instead provide the means of testing the {the theory outlined in Section \ref{SHM} }for realistic nonlinear {system-scale} scenarios of strong fast-forward IP shock-induced magnetopause motion. 
Large-scale periodic magnetopause motion, consistent with those described here, have been observed following the arrival of strong fast-forward IP shocks. During the impact of the August 1972 ICME, when the sub-solar magnetopause was compressed to less than 5.2 R$_E$ upstream, the Explorer 45 satellite experienced multiple magnetopause crossings in rapid succession \citep{Cahill77}. Similarly, during the extreme event of March 1991, the GOES-6 satellite experienced six inward-outward periodic movements of the magnetopause over a 30 minute period \citep{Cahill92}.  The lack of an upstream solar wind monitor does, however, complicate further direct comparison to these events.

\subsection{Frequency Analysis}

\begin{figure*}[ht]
\centering
\hspace*{-1cm}
\includegraphics[width=1\textwidth]{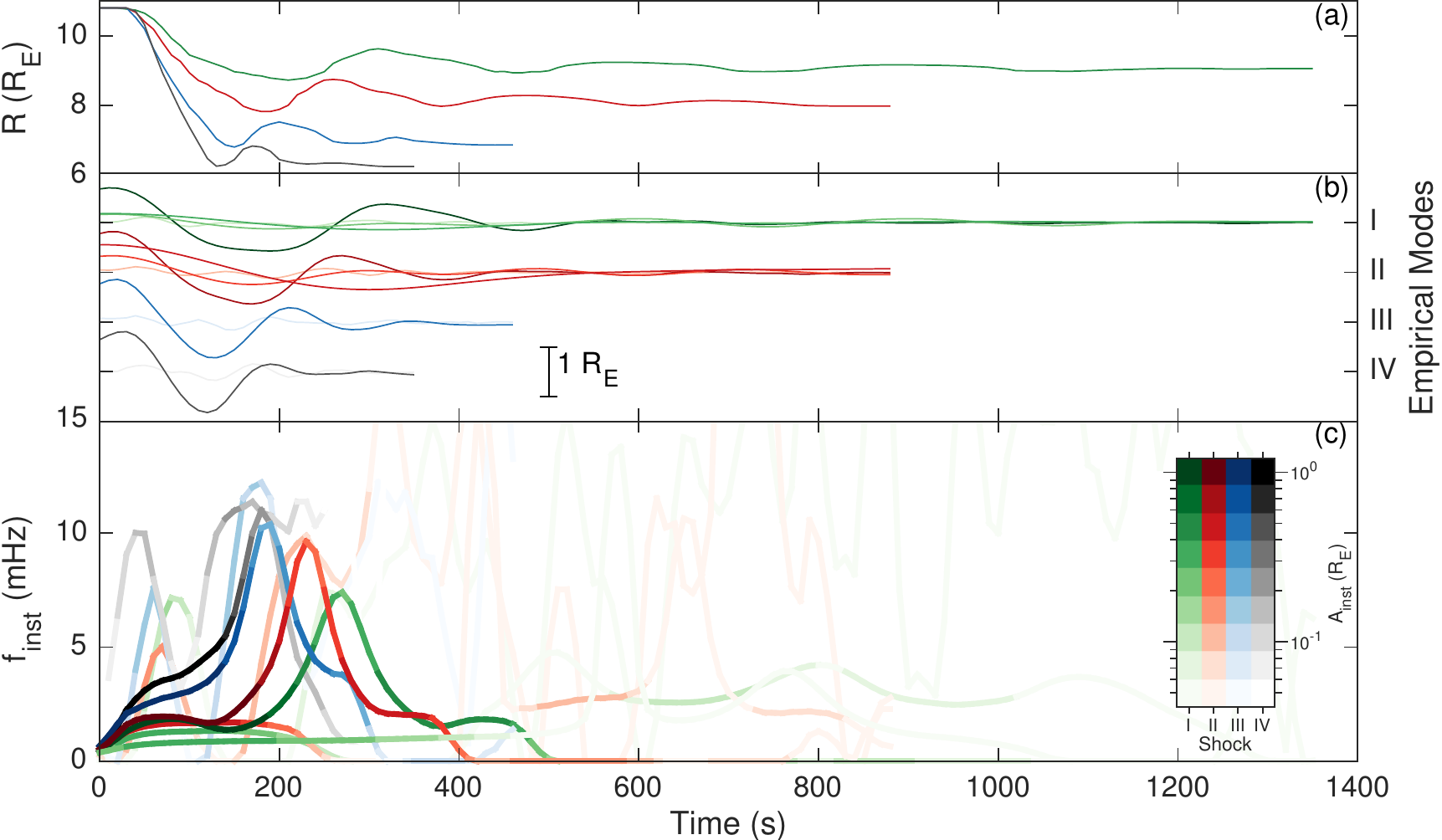}
\caption{(a) Shows the original modes shown in Figure \ref{subsolar}, (b) shows these decomposed into their empirical modes and (c) shows a Hilbert transform of their instantaneous frequencies.
\label{EEMD}}
\end{figure*}

The response of the magnetopause in the Gorgon simulations requires time-frequency analysis suitable for non-stationary and nonlinear processes. Figure \ref{EEMD} uses ensemble empirical mode decomposition (EEMD) \citep{Wu04,Torres11} to derive the statistically significant modes associated with the magnetopause motion and a Hilbert transform spectrum shows the associated characteristic frequencies.

These show the oscillations as a function of up to four statistically significant modes, the primary of which exhibit frequencies between 2--13 mHz with the frequencies of the dominant modes increasing with shock strength. 
The instantaneous frequencies initially increase from zero as the inertial phase begins and then plateaus somewhat during the compressive phase. They then rapidly increase during the first magnetopause rebound before relaxing back to values between 2--5 mHz. The instantaneous frequency is the time-derivative of the phase at each moment and this distinct peak is therefore interpretted as evidence of nonlinear phase steepening.
Due to the strong damping, the instantaneous frequencies don't always provide a good handle on the overall periodicity of the oscillations in each mode. Taking the auto-correlation of each mode and finding the peak, we find slightly higher overall frequencies of:
3.3, 4.1, 5.4 and 5.8 mHz for Shocks I--IV, respectively. 
 The primary empirical modes appear strongly damped with Shocks I and II inducing approximately three total periods of diminishing amplitude whereas Shocks III and IV induce less than two such periods.
The eventual periods of the oscillations are much longer than the first few oscillations for both Shocks I and II. These are being picked up by the secondary mode and the frequencies appear in the range of magnetopause surface eigenmodes being reflected between the northern and southern ionospheres \citep{Chen74} as seen in high-resolution global MHD simulations \citep{Hartinger15} at 1.8 and 2.3 mHz respectively. With Shocks III and IV these are not apparent, possibly due to the grid resolution not sufficiently resolving field-aligned currents near the inner boundary and magnetopause reconnection at the subsolar point from the strong southward driving prohibiting a surface eigenmode forming \citep{Plaschke11,Archer19}. Further modes are apparent extending up to 0.1 Hz but these likely correspond to nonlinear higher order terms.

{
The simulated magnetopause frequencies at the subsolar point lie where the natural frequencies of the magnetopause fall according to the theory outlined in Section \ref{SHM}. 
These oscillations notably occur at the lower end of the ULF range observed throughout the magnetosphere \citep{Menk2011} 
and 
\citet{Freeman1995} point out that the linear theory predicts that the magnetopause acts as a low pass filter of compressional waves due to solar wind dynamic pressure variations and resonances may thus be selectively enhanced at the natural eigenfrequency and suppressed at higher frequencies. 
Higher frequency waves, however, could well exist further within the magnetosphere, for example via field line resonances excited by the fast magnetosonic pulse during the compression phase.
The reproduction of ULF waves in global-MHD simulations can, however, be sensitive to numerical effects \citep{Claudepierre09} and an exploration of the magnetospheric ULF counterparts to IP shocks is therefore left for a future endeavour}.

\section{Conclusions}
\label{summary}

This study has examined the magnetospheric and magnetopause response to four synthetic IP shocks of varying magnitudes using Global-MHD simulations. While previous studies \citep{Smit1968,Freeman1995,Borve2011} focussed on small-scale dynamic pressure changes in the upstream driver, {we developed nonlinear theory suitable for large peturbations and compared these to} self consistent global MHD simulations. This approach enabled the characterisation of magnetopause motion for extreme scenarios representative of fast-forward shocks striking the magnetosphere, as occur at the forefront of coronal mass ejections.

In response to the IP shocks, the simulated magnetopause notably featured large-scale oscillatory motion  of the order of an Earth radius, prior to reaching pressure balance. This was readily explained when considering the driving, inertial and restoring forces associated with theory of the magnetopause as a forced damped simple harmonic oscillator. The frequencies of the oscillations occurred in the range of 2--13 mHz, predominantly occurring between 2--5 mHz. {The response times and oscillation periods seen in the simulations were quantitatively consistent with the nonlinear theory, and the damping time of the oscillation was also quantitatively consistent with nonlinear theory for the stronger shocks but underestimated by theory for the weaker shocks. The initial magnetopause response was also best predicted by linear theory for the weaker shocks and by nonlinear theory for the strongest shock, which is consistent with the assumptions beyond deriving the linearised solutions.}

These large-amplitude oscillations provide an explanation for periodic magnetopause motion observed following the impact of strong interplanetary shocks during the extreme space weather events of August 1972 \citep{Cahill77} and March 1991 \citep{Cahill92}. The time-delay in the magnetopause response due to the inertia of the magnetosheath, combined with the large-scale oscillatory motion, also helps to understand why static models of the magnetopause break down during periods of strong solar wind driving \citep[e.g.][]{Staples2020}. Furthermore, the varying structure throughout a given Earth-bound coronal mass ejection, combined with the dynamic magnetopause response, could well mean that the magnetopause rarely settles into highly compressed equilibrium states, which would also introduce a significant bias to in-situ measurements of its locations.

Rapid inward motion of the magnetopause has been observed to consistently produce enhancements and dropouts in the radiation belt phase space distributions \citep{Reeves03,Schiller16} and to drive an abundance of global ultra-low-frequency wave activity \citep{Li97,Green04} and enhance ionospheric and ground-induced currents \citep{Fujita03,Smith19}. These phenomena therefore present further observables which could be affected by and tested {\citep[e.g.][]{Wang10}} for the large-scale magnetopause oscillations described herein.

%%%%%%%%%%%%%%%%%%%%%%%%%%%%%%%%%%%%%%%

\section*{Acknowledgements}
 RTD, JPE and JPC acknowledge funding from NERC grant NE/P017347/1 (Rad-Sat). MPF was supported by NERC grant NE/P016693/1 (SWIGS). JWBE is funded by a UK Science and Technology Facilities Council (STFC) Studentship (ST/R504816/1). MOA holds a UKRI (STFC /
EPSRC) Stephen Hawking Fellowship EP/T01735X/1. Research into magnetospheric modelling at Imperial College London is also supported by Grant NE/P017142/1 (SWIGS). NM and RH would like to acknowledge the Natural Environment Research Council Highlight Topic grant NE/P10738X/1 (Rad-Sat) and the NERC grants NE/V00249X/1
298 (Sat-Risk) and NE/R016038/1. IJR and FAS acknowledge STFC grants ST/V006320/1 and NE/P017185/1. This project has received funding from the European Union's Horizon 2020 research and innovation programme under grant agreement No. 870452 (PAGER). This work used the Imperial College High Performance Computing Service (doi: 10.14469/hpc/2232).  
\newline

\section*{Data Availability Statement}
The simulation data used in this paper is openly available on the UK Polar Data Centre (UK PDC):  https://doi.org/10.5285/3774fa5b-f2fb-42c3-9091-5b11ac9744ea

\bibliographystyle{abbrvnat}
\bibliography{mybibfile}

\end{document}